\documentclass[12pt,preprint]{aastex}



\newcommand{\js}{}

\newcommand{\ie}{{\js i.e.\/}}

\lefthead{Huang et al.~}

\slugcomment{DRAFT \today}
\shorttitle{IRAC Imaging of Lockman Hole}
\shortauthors{Huang et al.}
\begin{document}
\title{IRAC Imaging of the Lockman Hole}
\author{J.-S.\ Huang,$\!$\altaffilmark{1}
P. Barmby,$\!$\altaffilmark{1}
G. G. Fazio,$\!$\altaffilmark{1}
S. P. Willner,$\!$\altaffilmark{1}
G. Wilson, $\!$\altaffilmark{2}
D. Rigopoulou, $\!$\altaffilmark{3}
A. Alonso-Herrero,$\!$\altaffilmark{4}
H. Dole,$\!$\altaffilmark{4,}$\!$\altaffilmark{5}
E. Egami,$\!$\altaffilmark{4}
E. Le Floc'h,$\!$\altaffilmark{4}
C. Papovich,$\!$\altaffilmark{4}
P.~G.  P\'erez-Gonz\'alez,$\!$\altaffilmark{4}
J. Rigby,$\!$\altaffilmark{4}
C. W. Engelbracht,$\!$\altaffilmark{4}
K. Gordon,$\!$\altaffilmark{4}
D. Hines,$\!$\altaffilmark{4}
M. Rieke,$\!$\altaffilmark{4} 
G. H. Rieke,$\!$\altaffilmark{4}
K. Meisenheimer$\!$\altaffilmark{6}
S. Miyazaki$\!$\altaffilmark{7}
}

\altaffiltext{1}{Harvard-Smithsonian Center for Astrophysics, 60 Garden Street,
Cambridge, MA02138}
\altaffiltext{2}{Spitzer Science Center,
Caltech, 1200 E. California, Pasadena, CA91125}
\altaffiltext{3}{Department of Astrophysics, Oxford University, Keble Rd, Oxford, OX1 3RH, U.K.}
\altaffiltext{4}{Steward Observatory, University of Arizona, Tucson, AZ85721}
\altaffiltext{5}{Institut d'Astrophysique Spatiale, bat 121, 
Universite Paris Sud, F-91405 Orsay Cedex, France}
\altaffiltext{6}{Max-Planck-Institut f\"{u}r Astronomie, K\"{o}nigstuhl 17,
D-69117 Heidelberg, Germany}
\altaffiltext{7}{Subaru Telescope,National Astronomical Observatory of Japan, 
                    650 North A'ohoku Place,
                    Hilo, HI 96720}

\slugcomment{Accepted by the ApJS, Spitzer Special Issue}

\begin{abstract}

IRAC imaging of a  
$4\farcm7\times4\farcm7$ area in the Lockman Hole detected over
400 galaxies in the IRAC 3.6$\mu$m 
and 4.5$\mu$m bands, 120 in the  5.8$\mu$m band, and 80 in the
8.0$\mu$m band in 30 minutes of observing time.
Color-color diagrams suggest that about half of these galaxies are at 
redshifts $0.6<z<1.3$ with about a quarter
at higher redshifts ($z>$1.3).
We also detect IRAC counterparts for 6 of the 7 SCUBA sources 
and all 9 XMM sources in this area. The detection
of the counterparts of the SCUBA sources and galaxies at $z>1.3$
demonstrates the ability
of IRAC to probe the universe at very high redshifts.

\end{abstract}
\keywords{cosmology: observations --- galaxies: evolution --- galaxies: 
Survey --- galaxies: Mid-Infrared}


\section{Introduction}

Much effort has gone into deep optical imaging with hope of detecting
galaxies at high redshifts \citep{koo92, ell97}.  Obtaining redshifts
for faint, optically selected galaxies require long observations on
$\ge$8-meter class telescopes, but most of the galaxies identified
are at intermediate redshifts (z$\sim$1) \citep{cow96,coh99,cim02}.
Applying faintness criteria to visible-wavelength (UBVRI) 
surveys is not an efficient way of selecting high-redshift galaxies
unless color
selection methods such as the Lyman break technique \citep{ste03} or
extreme red color selection \citep{tho99} are applied.  Furthermore,
optical studies of high redshift galaxies are heavily affected by
dust extinction because rest-frame UV photons are being observed
\citep{ste03}.

By contrast, far-infrared and submillimeter surveys are
efficient tools for generating samples of high-redshift starburst galaxies.
For example, SCUBA on the JCMT
has made it possible to carry out deep extragalactic surveys in the
submm band. Now it is known that most SCUBA sources are ultra luminous
infrared galaxies (ULIRG) at high redshifts. 
They are detected in the submillimeter band,
because the infrared luminosity peak at about 100--200 \micron\ shifts
to the submillimeter and millimeter band at z$>$2 \citep{cha03,ivi02,sma02}.

Little is known about the properties of mid-infrared selected
samples, although in principle they should have many advantages.  
Mid-infrared observations of high-redshift galaxies detect light 
emitted at 1-2 \micron\, near the peak of the stellar emission.  
At these wavelengths the emission is only modestly
affected by dust extinction. The
difficulty is that a space observatory is essential for sensitive
mid-infrared observations.  The Infrared Space Observatory (ISO)
\citep{kes96} carried out the first deep infrared extragalactic
surveys and detected galaxies at redshifts up to $z=1.5$ in the
mid--infrared bands \citep{oli97, elb99, aus99, gen00, dol01, sat03}.
With its combination of high sensitivity and high spatial resolution in four
mid-infrared bands, the Infrared Array Camera (IRAC) \citep{faz04} on
the Spitzer Space Telescope (SST) can carry out deeper and larger
surveys and should be able to detect galaxies at
$z>3$ \citep{bar04}.
 
This Letter reports the results of a SST survey carried out during the
SST In-Orbit-Checkout (IOC) period. Observations were taken as one of
the Early Release Observations to demonstrate the sensitivity
of Spitzer and its capability of detecting galaxies at high
redshifts. The program included both IRAC and MIPS
observations of the same Lockman Hole field.
\citet{ega04} and \citet{emr04}  present the
results of the MIPS observations of the same field.  
The survey area centered on RA(J2000)=10:51:56, DEC(J2000)=57:25:32
has been covered
by surveys at many other wavelengths, including SCUBA, XMM, and 
deep optical/NIR imaging.  Of seven SCUBA sources in 
the survey area, IRAC detected six in the 3.6
and 4.5~\micron\ bands and four in the 5.8 and 8.0~$\mu$m bands.  IRAC
also detected mid-infrared counterparts of all 9 XMM sources.
Analysis of the Spitzer counterparts of the SCUBA, XMM and MAMBO
sources are presented by \citet{ega04}, \citet{alo04} and \citet{ivi04}.  


\section{Observations and data reduction}

The observations\footnote{Spitzer Program ID 1077, executed 2003 November 20}
covered one field-of-view in all four IRAC bands. 
The observations used the 30-s frame time and 25 dithers per
pointing, giving an exposure time (\ie, effective on-chip charge
accumulation time after accounting for readout overheads) of 670~s in
each band.  We used a
shorter frame time and many dithers to be able to reject cosmic-rays
effectively.  Because of the dithering, the area covered to full
depth was $4\farcm7 \times 4\farcm7$.
 
Data processing began with the Basic Calibrated Data (BCD)
delivered by the Spitzer Science Center (SSC).  
These data include flat-field corrections,
dark subtraction, and linearity and flux calibrations.  Additional
steps included pointing refinement, distortion correction, and
mosaicking. Cosmic rays were rejected during the last step by
sigma-clipping. 

Source extraction used DAOPHOT because the images are so crowded
in the 3.6$\mu$m and 4.5$\mu$m bands even for the short
observations reported here.  Given the 1\farcs8 FWHM of the point
spread function \citep{faz04} in both bands, almost all faint sources
are point-like.  We used a signal/noise ratio of 2.5, and a PSF with
full-widths at half-maximum (FHWM) of 1\farcs8, 1\farcs8, 2\farcs0, 
and 2\farcs0 in the 3.6, 4.5, 5.8, and 8.0~\micron\  bands respectively
as parameters for DAOPHOT.

Star/galaxy separation was based on morphology in deep R-band images
taken with the Suprime camera on the Subaru Telescope and on a ($V-I$
vs $I-K$) color-color diagram \citep{hua97,wil01}.  We identified 13
stars in the field with $K\le20.5$. Though fainter stars may be
missed, their density is extremely low \citep{gla94}.
Table~1 gives the numbers of galaxies and
stars detected in each band along with the limiting flux.
The number of galaxies detected in the MIPS 24$\mu$m band \citep{ega04, emr04}
is also in Table~1.
Fewer sources were detected in the two long wavelength bands
because of the high zodiacal background at those wavelengths.  All
objects detected at 5.8 or 8.0~\micron\ are detected at both 3.6 and
4.5~\micron\.  There are about 25000 ``beams'' in the field, and the
3.6~\micron\ source counts thus correspond to 57 beams per source.

\section{The Properties of Galaxies Detected by IRAC}

With no redshifts available for most of the IRAC galaxy sample, we
rely on color-color diagrams to estimate redshifts.  In the rest 
near-infrared wavelength range, the most important spectral feature is the
1.6~$\mu$m bump, which is caused by the H$^{-}$ opacity minimum 
in the atmospheres of cool stars. This feature is seen in 
the spectral energy distributions (SEDs)
of all galaxies except AGN-dominated ones
and has long been considered a photometric redshift indicator
\citep{wri94, sim99, saw02}.  At $z > 0.6$, the bump shifts beyond the
$K$ band, and $f_{\nu}(3.6\micron) > f_{\nu}(2.2\micron)$. For
$z>1.3$, the bump passes 3.6~\micron, and $f_{\nu}(4.5\micron)
> f_{\nu}(3.6\micron)$.   Local, dusty starburst 
galaxies such as M82
also have red colors with $f_{\nu}(4.5\micron) >
f_{\nu}(3.6\micron)$, and we need to require $f_{\nu}(3.6\micron) >
f_{\nu}(2.2\micron)$ to identify them. 

Figures~1 and~2 show the IRAC data along with the tracks followed by
typical galaxy SEDs as redshift increases,
and Table~2 summarizes the observed galaxy number distribution as a function
of color, which imply redshift.  Figure~1 shows that the observed
colors are consistent with the model tracks. The most-populated
redshift bin in Table~2 for all four IRAC bands is $0.6 < z < 1.3$.
This bin includes about 50\% of all detected IRAC galaxies. 
Figure~1 shows a few sources with very red K-[3.6] color.
\citet{wil04} found that these sources are actually 
the same extreme red objects as those detected in optical-near-infrared
surveys using the (R-K)$>$5 color selection.
 
The color-color diagrams will not give accurate redshifts for AGN,
which do not exhibit the 1.6$\mu$m bump in their SEDs. X-ray
observation is a good way of detecting AGN. This field has 9 XMM
sources \citep{has01}, whose colors are plotted along with other
galaxies in Figures~1 and~2.  These 9 sources probably represent
most of the AGN in the field and constitute only a few percent of the
galaxies detected.  Thus they should not affect our statistics on
redshift distribution.

Unlike the XMM sources, whose color distribution fills the entire
color space in Figures~1 and~2, most of the SCUBA source counterparts have
colors indicating $z>1.3$.  Using the deep radio counterpart positions to
identify the SCUBA sources in the optical and Spitzer images, 
\citet{ega04} did a detailed analysis of
SED for the SCUBA sources. They found that LE~850.1\footnote{The full names 
according to SINBAD are of the form [SFD2002]LHE~N.}, 
LE~850.4, LE~850.7, 
and LE~850.14 have the 1.6\micron~ bump 
in their SEDs.
This implies that
their SEDs are dominated by stellar light, and thus
their IRAC colors following the galaxy color tracks well in Figures~1 and~2
with expected redshfits $z>2$.
This is consistent with most of the SCUBA sources
being dusty galaxies at $z>2$ \citep{cha03, ega04}. 
LE~850.8 is the only object for which
we have both SCUBA and XMM detection. \citet{ega04} found that LE~850.8  has a
power-law continuum.
Its IRAC colors are not close to any of the tracks in Figures~1 and~2.
Another SCUBA source, LE~850.18,
also has colors not close to any of the tracks
and a power-law SED \citep{ega04}.
In the deep R-band image, however, there are more than 2 objects within
$3''$ radius centered on the radio counterpart positions for both 
LE~850.8 and LE~850.18. IRAC will not be able to resolve them if all components
emit in the IRAC bands.
Hence we suspect foreground contamination in the IRAC flux
densities in both cases and are cautious in using IRAC colors to classify
both objects.

There are 32 sources in this field detected in the MIPS
24~\micron\ band at or above the 5-sigma confidence level.
Their IRAC counterparts show that half of the
MIPS sources are at $z>1.3$. Given the 5$\sigma$ limiting flux of 200
$\mu$Jy, these MIPS sources at $z>1.3$ have a minimun infrared
luminosity of 10$^{12}$ L$_{\odot}$, thus are LIRGs and ULIRGs \citep{emr04}.

The extragalactic background light (EBL) from both resolevd and 
unresolved extragalactic sources is an indicator of the total luminosity of
the Unverse \citep{mad00}.
Because the intermediate
redshift galaxy population contributes most to the EBL \citep{elb02}, 
integral resolved galaxy light (IGL) in a deep survey
should give a good lower limit on the EBL. 
Table~3 shows the results of
integrating the fluxes, $\nu I_{\nu}$, from IRAC counterparts
of objects identified by
SCUBA \citep{sco01}, XMM \citep{has01}, and MIPS 24~\micron\ images
\citep{ega04, emr04}.  
The IGL at the 5.8~\micron\ and 8.0~\micron\ 
are 1.31 and 1.02 nW m$^{-2}$ sr$^{-1}$, very
close to the ISO LW2 (6.75~\micron) IGL, 1.7$\pm$0.5 nW m$^{-2}$ sr$^{-1}$
\citep{alt99, elb02}.
Given that the XMM sources in
this field are neither stars or clusters, it is safe to assume that
most of them are AGN \citep{leh00,alo04}. The light
contributed by the XMM sources is constant in all four IRAC bands
because of the flat SED of AGN in the mid-infrared.  Because the
total light decreases with wavelength, the XMM
source contribution increases from 3.7\% at 3.6~$\mu$m to 13.6\% at
8~$\mu$m. This is consistent with the 15$\pm$5\% XMM source
contribution found by the ISO survey \citep{fad02}. As shown in
Figure~2, most of the MIPS 24~\micron\ sources are at
intermediate and high redshifts.  Most of them must be LIRG's or even
ULIRG's \citep{ega04}.  The starburst light percentage increases from
20\% at 3.6~$\mu$m to 62\% at 8~$\mu$m. AGN and starburst galaxies
together contribute about two thirds of the IGL at 8~$\mu$m.  The bulk
of the light at 3.6~$\mu$m and 4.5~$\mu$m appears to 
come from evolved stars in galaxies, while the 8~$\mu$m and 24~$\mu$m
light traces the star forming regions in galaxies. This explains
the large contribution of the 24$\mu$m selected sample to the 8$\mu$m
IGL.

\section{CONCLUSIONS}

Even with only 30 minutes observing time, IRAC detected
419 galaxies in a $4\farcm7\times4\farcm7$ area in the Lockman Hole.
Based on their colors, most of the detected galaxies are likely to be
at intermediate redshifts $0.6<z<1.3$, but $\ga$25\% appear to be at
$z>1.3$.   30\% of the 5.8~$\mu$m sources and 50\% of the
8.0~$\mu$m sources show colors indicating higher redshifts (z$>$1.3).  
The counterparts of the SCUBA sources have colors
consistent with galaxies at $z>1.3$.  The
integrated galaxy light seen in our sample is consistent with
1.7$\pm$0.5 nW m$^{-2}$ sr$^{-1}$ derived using the ISO survey 
data at the 6.75~$\mu$m.  
The AGN fraction, as revealed by XMM, increases from
3.7\% in the 3.6~$\mu$m band to 13.6\% in the 8.0~$\mu$m band.  The MIPS
population, most of which are starburst galaxies at intermediate
redshifts, contributes up to 60\% in the 8.0$\mu$m band.

\acknowledgements

This work is based on observations made with the Spitzer Space
Telescope, which is operated by the Jet Propulsion Laboratory,
California Institute of Technology under NASA contract 1407. Support
for this work was provided by NASA through Contract Number 1256790
issued by JPL.



\clearpage


\clearpage
\begin{figure}
\plotone{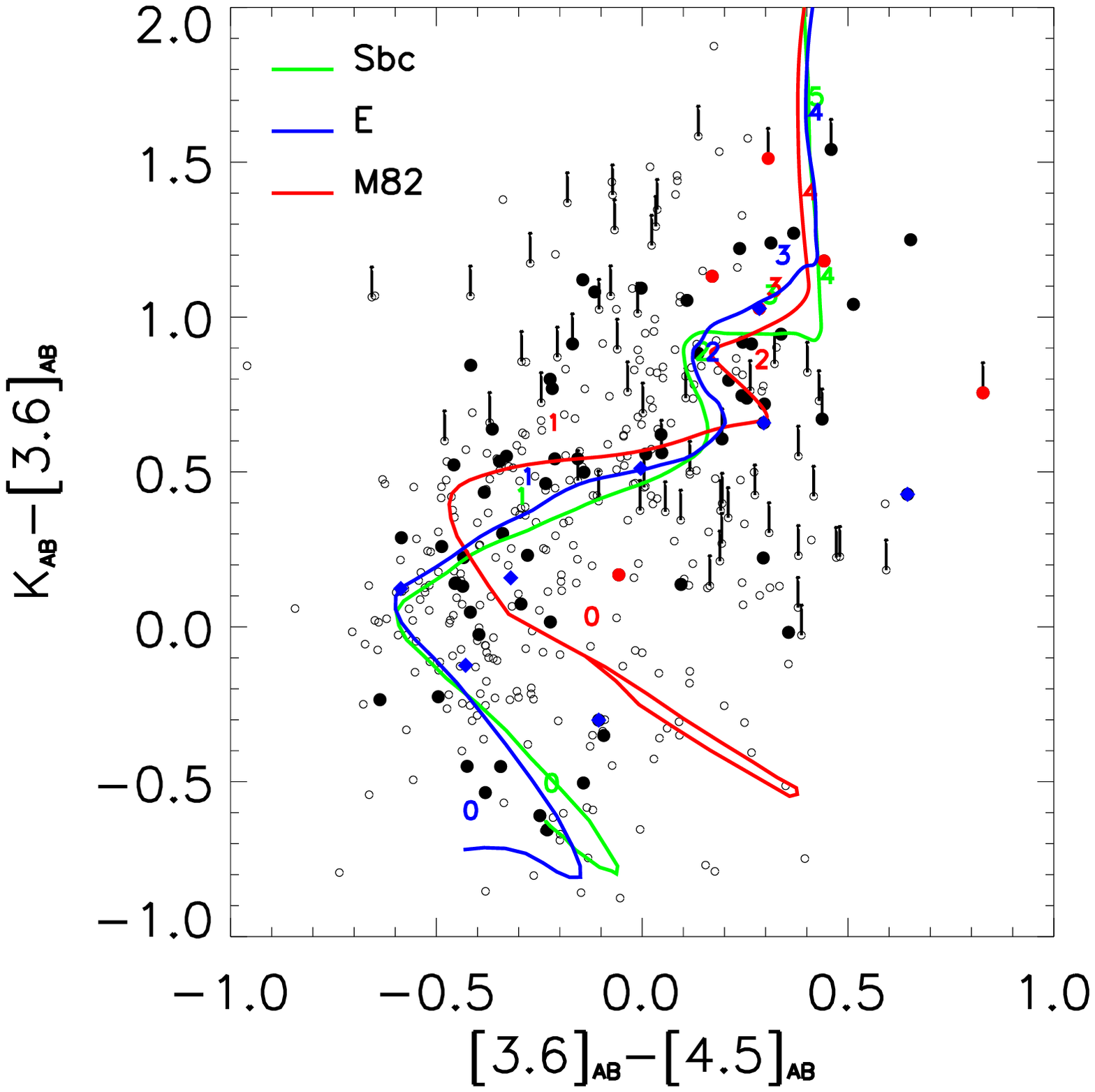}
\caption{Color-color diagram for objects detected at 3.6~$\mu$m and
4.5~$\mu$m.  Filled circles are objects also detected at 
24~$\mu$m, open circles are remaining sample.  
Arrows indicate objects not detected
at $K$, points are plotted at 3$\sigma$ limits. Blue diamonds
denote XMM sources, and red circles denote SCUBA sources.  
Solid lines show color tracks for
elliptical galaxies (blue), Sbc galaxies (green), and M82 (red).
Expected colors for ellipticals and spirals are based on rest frame
SED's from \citet{col80} in visible light and \citet{lun03} in the
infrared.  M82 tracks are derived from the model of \citet{gra00} and
ISO observations \citep{stu00}.  The tracks are derived by shifting
the SED without adding any evolution; numbers marked along each track
indicate redshift.  In this diagram, galaxies in the upper left
quadrant are likely to be at $0.6<z<1.3$, those in the upper right
are likely to be at $z>1.3$, and galaxies in the two lower quadrants are
likely to have $z<0.6$ with starburst galaxies to the right.
\label{fig1}}
\end{figure}

\clearpage
\begin{figure}
\plotone{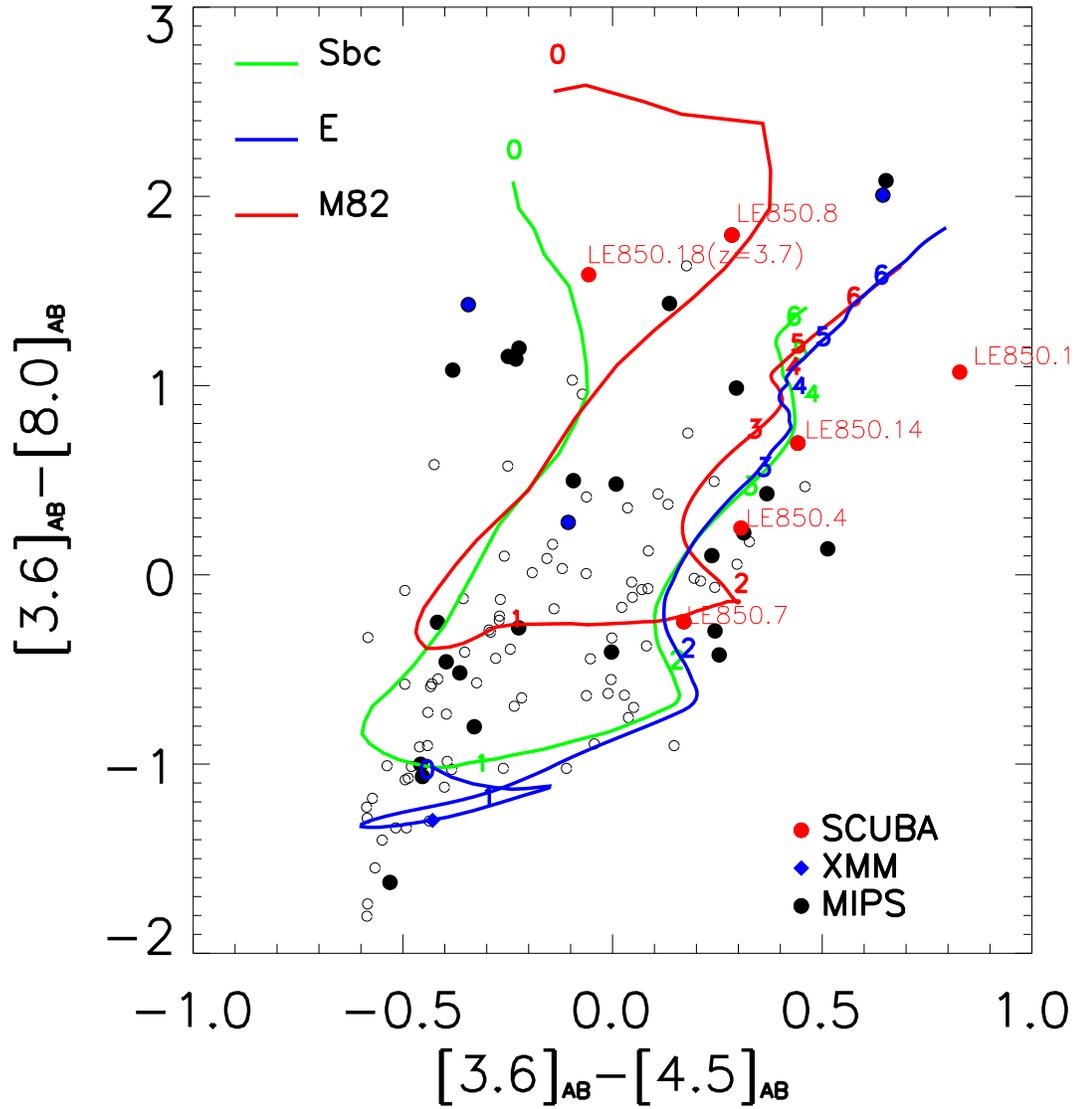}
\caption{
Color-color diagram for galaxies detected at 8~\micron. Blue diamonds
show XMM sources, red dots SCUBA sources with numbers according to
\citet{sco01}, and black dots are
objects detected at 24~$\mu$m. 
Open circles show the remaining sample galaxies. 
Color tracks are the same as those in Figure~2.
\label{fig2}
}
\end{figure}

\clearpage

\begin{table*}[hbt]
{\scriptsize
\begin{center}
\centerline{\sc Table 1}
\vspace{0.1cm}
\centerline{\sc Number of Objects detected in the IRAC bands}
\vspace{0.3cm}
\begin{tabular}{lcccc}
\hline\hline
\noalign{\smallskip}
 Band & Galaxies & Stars & Total & Mag Limit(5$\sigma$,AB)\cr
\hline
\noalign{\smallskip}
3.6$\mu$m & 419 & 13 & 432 &23.73\cr
4.5$\mu$m & 403 & 13 & 413 &23.77\cr
5.8$\mu$m & 120 & 5 & 125  &21.90\cr
8.0$\mu$m &  80 & 5 & 85   &21.68\cr
24.0$\mu$m & 32 & 0 & 32   &18.15\cr
\noalign{\hrule}
\noalign{\smallskip}
\end{tabular}

\end{center}
}
\label{tab1}
\end{table*}

\begin{table*}[hbt]
{\scriptsize
\begin{center}
\centerline{\sc Table 2}
\vspace{0.1cm}
\centerline{\sc Galaxies in each IRAC Color Bin }
\vspace{0.3cm}
\begin{tabular}{cccccccc}
\hline\hline
\noalign{\smallskip}
$K_{AB}-[3.6]_{AB}$ & $[3.6]_{AB}-[4.5]_{AB}$  & Estimated Redshift & 3.6$\mu$m &4.5$\mu$m &5.8$\mu$m&8.0$\mu$m&24$\mu$m\cr
\hline
\noalign{\smallskip}
$>0$& $>0$ & $z>1.3$& 113&122&35&24&16\cr
$>0$ & $<0$&$0.6<z<1.3$&191&176&57&34&8\cr
$<0$& $<0$ &$z<0.6$ & 101 &81&27&22&8\cr
$<0$& $>0$ &$z<0.6$& 14 & 24&1&0&0\cr
\noalign{\hrule}
\noalign{\smallskip}
\end{tabular}

\end{center}
}
\label{tab2}
\end{table*}

\begin{table*}[hbt]
{\scriptsize
\begin{center}
\centerline{\sc Table 3}
\vspace{0.1cm}
\centerline{\sc Integrated Galaxy Light $\nu$ I$_{\nu}$ (nW m$^{-2}$ sr$^{-1}$) in the IRAC Bands }
\vspace{0.3cm}
\begin{tabular}{lcccc}
\hline\hline
\noalign{\smallskip}
 ~~& 3.6$\mu$m &4.5$\mu$m &5.8$\mu$m&8.0$\mu$m\cr
\hline
\noalign{\smallskip}
All galaxies detected& 3.47 & 2.43 & 1.31& 1.03\cr
SCUBA counterparts & 0.05 & 0.05 & 0.05 &0.04\cr
XMM counterparts& 0.13& 0.12 & 0.13 & 0.14\cr
MIPS counterparts & 0.68 & 0.52 & 0.43 & 0.62\cr
\noalign{\hrule}
\noalign{\smallskip}
\end{tabular}
\end{center}
}
\label{tab3}
\end{table*}

\end{document}